\documentclass[sigconf]{acmart}

\def\BibTeX{{\rm B\kern-.05em{\sc i\kern-.025em b}\kern-.08emT\kern-.1667em\lower.7ex\hbox{E}\kern-.125emX}}
    
\usepackage{acronym}
\newacro{sme}[SME]{Small and Medium-sized Enterprise}
\newacro{it}[IT]{Information Technology}
\newacro{ot}[OT]{Operation Technology}
\newacro{cots}[COTS]{Commercial Off-The-Shelf}
\newacro{iot}[IoT]{Internet of Things}
\newacro{plc}[PLC]{Programmable Logic Controller}
\newacro{cps}[CPS]{Cyber-Physical System}
\newacro{cpps}[CPPS]{Cyber-Physical Production System}
\newacro{ids}[IDS]{Intrusion Detection System}
\newacro{svm}[\textit{SVM}]{\textit{Support Vector Machine}}
\newacro{wsn}[WSN]{Wireless Sensor Network}
\newacro{darpa}[DARPA]{Defense Advanced Research Projects Agency}
\newacro{kdd}[KDD]{Knowledge Discovery in Databases}
\newacro{scada}[SCADA]{Supervisory Control And Data Acquisition}
\newacro{dpi}[DPI]{Deep Packet Inspection}
\newacro{rnn}[RNN]{Recurrent Neural Network}
\newacro{s317}[\textit{S317}]{\textit{SUTD Security Showdown (S3) 2017}}
\newacro{dmz}[DMZ]{De-Militarized Zone}
\newacro{svm}[\textit{SVM}]{\textit{Support Vector Machine}}
\newacro{ocsvm}[\textit{OCSVM}]{\textit{One-Class Support Vector Machine}}
\newacro{lstm}[\textit{LSTM}]{\textit{Long Short-Term Memory}}
\newacro{iiot}[IIoT]{Industrial Internet of Things}
\newacro{opcua}[\textit{OPC UA}]{\textit{Object Linking and Embedding for Process Control Unified Architecture}}
\newacro{swat}[\textit{SWaT}]{\textit{Secure Water Treatment}}
\newacro{pca}[\textit{PCA}]{\textit{Principal Component Analysis}}
\newacro{sssp}[SSSP]{Single Stage Single Point}
\newacro{ssmp}[SSMP]{Single Stage Multi Point}
\newacro{mssp}[MSSP]{Multi Stage Single Point}
\newacro{msmp}[MSMP]{Multi Stage Multi Point}
\newacro{siem}[SIEM]{Security Information and Event Management}
\newacro{mtu}[MTU]{Master Terminal Unit}
\newacro{uf}[UF]{Ultra Filtration}
\newacro{ics}[ICS]{Industrial Control System}	
\newacro{uv}[UV]{Ultraviolet}    
\newacro{ro}[RO]{Reverse Osmosis} 		
\newacro{rbf}[RBF]{Radial Basis Function} 
\newacro{ics}[ICS]{Industrial Control System}

\usepackage{amsmath}	
\usepackage{mathtools}

\usepackage{multirow}

\usepackage{numprint}

\setcopyright{acmlicensed}
\copyrightyear{2019}
\acmYear{2019}
\acmConference[CECC 2019]{Central European Cybersecurity Conference}{November 14--15, 2019}{Munich, Germany}
\acmBooktitle{Central European Cybersecurity Conference (CECC 2019), November 14--15, 2019, Munich, Germany}
\acmPrice{15.00}
\acmDOI{10.1145/3360664.3360667}
\acmISBN{978-1-4503-7296-1/19/11}

\begin{document}

%
\title[Side Channel-aided IDS]{Discussing the Feasibility of Acoustic Sensors for Side Channel-aided Industrial Intrusion Detection: An Essay}

%
\author{Simon D. Duque Anton}
\email{simon.duque_anton@dfki.de}
\orcid{0000-0003-4005-9165}
\affiliation{%
  \institution{German Research Center for AI\\Intelligent Networks Research Group}
  \streetaddress{Trippstadter Str. 122}
  \city{Kaiserslautern}
  \country{Germany}
  \postcode{67633}
}

\author{Anna Pia Lohfink}
\email{lohfink@cs.uni-kl.de}
\affiliation{
  \institution{University of Kaiserslautern\\Department of Computer Science}
  \streetaddress{Trippstadter Str. 122}
  \city{Kaiserslautern}
  \country{Germany}
  \postcode{67633}
}

\author{Hans Dieter Schotten}
\email{hans_dieter.schotten@dfki.de}
\affiliation{
  \institution{German Research Center for AI\\Intelligent Networks Research Group}
  \streetaddress{Trippstadter Str. 122}
  \city{Kaiserslautern}
  \country{Germany}
  \postcode{67633}
}

%
\renewcommand{\shortauthors}{S. D. Duque Anton et al.}

%
\begin{abstract}
The fourth industrial revolution leads to an increased use of embedded computation and intercommunication in an industrial environment.
While reducing cost and effort for set up,
operation and maintenance,
and increasing the time to operation or market respectively as well as the efficiency,
this also increases the attack surface of enterprises.
Industrial enterprises have become targets of cyber criminals in the last decade,
reasons being espionage but also politically motivated.
Infamous attack campaigns as well as easily available malware that hits industry in an unprepared state create a large threat landscape.
As industrial systems often operate for many decades and are difficult or impossible to upgrade in terms of security,
legacy-compatible industrial security solutions are necessary in order to create a security parameter.
One plausible approach in industry is the implementation and employment of side-channel sensors.
Combining readily available sensor data from different sources via different channels can provide an enhanced insight about the security state.
In this work,
a data set of an experimental industrial set up containing side channel sensors is discussed conceptually and insights are derived.
\end{abstract}

%
%


%

%
\maketitle

\section{Introduction}
As attacks on industry have increased in effect and frequency over the last decades~\cite{Duque_Anton.2017a},
securing industrial networks an applications becomes crucial for industrial enterprises.
Since many field-bus and industrial Ethernet protocols do not contain means to ensure authentication and encryption,
gaining access to a network is sufficient for an attacker to read and participate on the communication.
Novel protocols such as \ac{opcua}~\cite{OPCFoundation.2017} provide means for security of communication,
but older protocols such as \textit{Modbus}~\cite{Modbus.2012, ModbusIDA.2006} and \textit{PROFINET}~\cite{PROFIBUS.2017} are still common in industrial \ac{ot} networks.
Due to the fourth industrial revolution,
communication and embedded computation devices are increasingly integrated into industrial environments.
Aside from the benefits in terms of reduced operation, set up and maintenance cost and effort,
these devices also increase the attack surface.
Historically,
two general assumptions motivated the reluctance to employ secure protocols~\cite{Igure.2006}:
Industrial networks being physically separated from public networks and the unique and application specific nature of industrial control networks that is infeasible for an attacker to comprehend.
The first assumption is broken by the fourth industrial revolution that relies heavily on intercommunication through network boundaries.
\ac{cots} hard- and software that makes integration, 
extension and set up easier enables attackers to gain intelligence about their targets as well,
breaking the second assumption.
Several recent attacks on industrial enterprises show the effects a cyber attack can have on a production environment once security measures are broken.
A relevant factor are the operation times of industrial production machines which are several decades.
Often, 
those machinery is not easily updated so that insecure legacy systems are operated.
In this work
an industrial data set with side-channel sensors is presented and analysed with respect to detectability of attacks in a qualitative fashion.
An overview of the state of the art is discussed in Section~\ref{sec:sota}.
The data set is presented in Section~\ref{sec:data},
the sensor data is evaluated in Section~\ref{sec:side_channels}.
Possibilities for intrusion detection are discussed in Section~\ref{sec:sc-based_id}.
A conclusion is drawn in Section~\ref{sec:conclusion}.

\section{Related Work}
\label{sec:sota}
There is considerable research about industrial cyber attacks.
Several white papers address the widely known successful attacks on industrial environments.
\textit{Stuxnet},
being the most famous,
has been discussed extensively~\cite{Dragos.2016, Langner.2013, Virvilis.2013, Lindsay.2013}.
However,
there has been a number of different attacks with similar goals and impacts,
such as \textit{Duqu}~\cite{Virvilis.2013},
\textit{Industroyer/Crashoverride}~\cite{Cherepanov.2017, Dragos.2016},
\textit{Flame}~\cite{Virvilis.2013},
\textit{BlackEnergy}~\cite{Cherepanov.2017, Dragos.2016},
\textit{Havex}~\cite{Dragos.2016} and \textit{Red October}~\cite{Virvilis.2013}.
A review of \ac{it} and \ac{ot} security of industrial enterprises by \textit{Positive Technologies} revealed many exploitable flaws~\cite{Positive_Technologies.2018}. \\ \par
In addition to work analysing the specific attacks,
research has been done to address the lack of security in industrial applications and protocols.
They provide scientific analyses of individual aspects of protocols commonly found in industrial environments.
\textit{Giehl et al.} provide a framework to assess security controls in manufacturing environments~\cite{Giehl.2019}.
\textit{Cherdantseva et al.} provide a survey of existing risk assessment methods and evaluate their usefulness with respect to \ac{scada} scenarios~\cite{Cherdantseva.2016}.
The detection of cyber attacks is discussed by \textit{Gao and Morris}~\cite{Gao.2014}.
Their main focus is \textit{Modbus}.
Attacks based on \textit{Modbus} environments are grouped into different classes.
Additionally,
they survey attacks on \acp{ics}~\cite{Morris.2013}.
\textit{Zhu et al.} take a similar approach and evaluate an abundance of different dimensions of consideration in industrial attacks~\cite{Zhu.2011}.
\ac{it} systems are systematically compared to \ac{ot} system,
among others with respect to the security objectives that are most important in the respective environment.
A taxonomy for \ac{scada}-specific attacks is presented by \textit{Zhu and Sastry}~\cite{Zhu.2010}.
Concepts for secure \ac{scada} systems are designed by \textit{Fernandez et al.} while presenting capabilities to evaluate said systems in terms of security~\cite{Fernandez.2010}. \\ \par
In addition to the industrial intrusion detection based on protocols and models,
there is a selection of works addressing side channels.
Commonly,
side channels are an attack vector found in various application areas,
such as acoustic side channel attacks on industrial manufacturing~\cite{AlFaruque.2016},
attacks on cryptographic hardware~\cite{Zhou.2005} and quantum key distribution~\cite{Zhao.2008}.
However,
\textit{Van Aubel et al.} present a method to employ side channels for intrusion detection in \acp{ics}~\cite{Van_Aubel.2017}.
They use the electro-magnetic field of a \ac{plc} during operation,
create motifs and compare those motifs to the motifs of a \ac{plc} under attack.
The electro-magnetic field is used by \textit{Strobel et al.} as well in order to disassemble firmware and detect attacks on the hardware level~\cite{Strobel.2015}.

\section{Data Set}
\label{sec:data}
\begin{figure}[h!]
\centering
\includegraphics[width=0.48\textwidth]{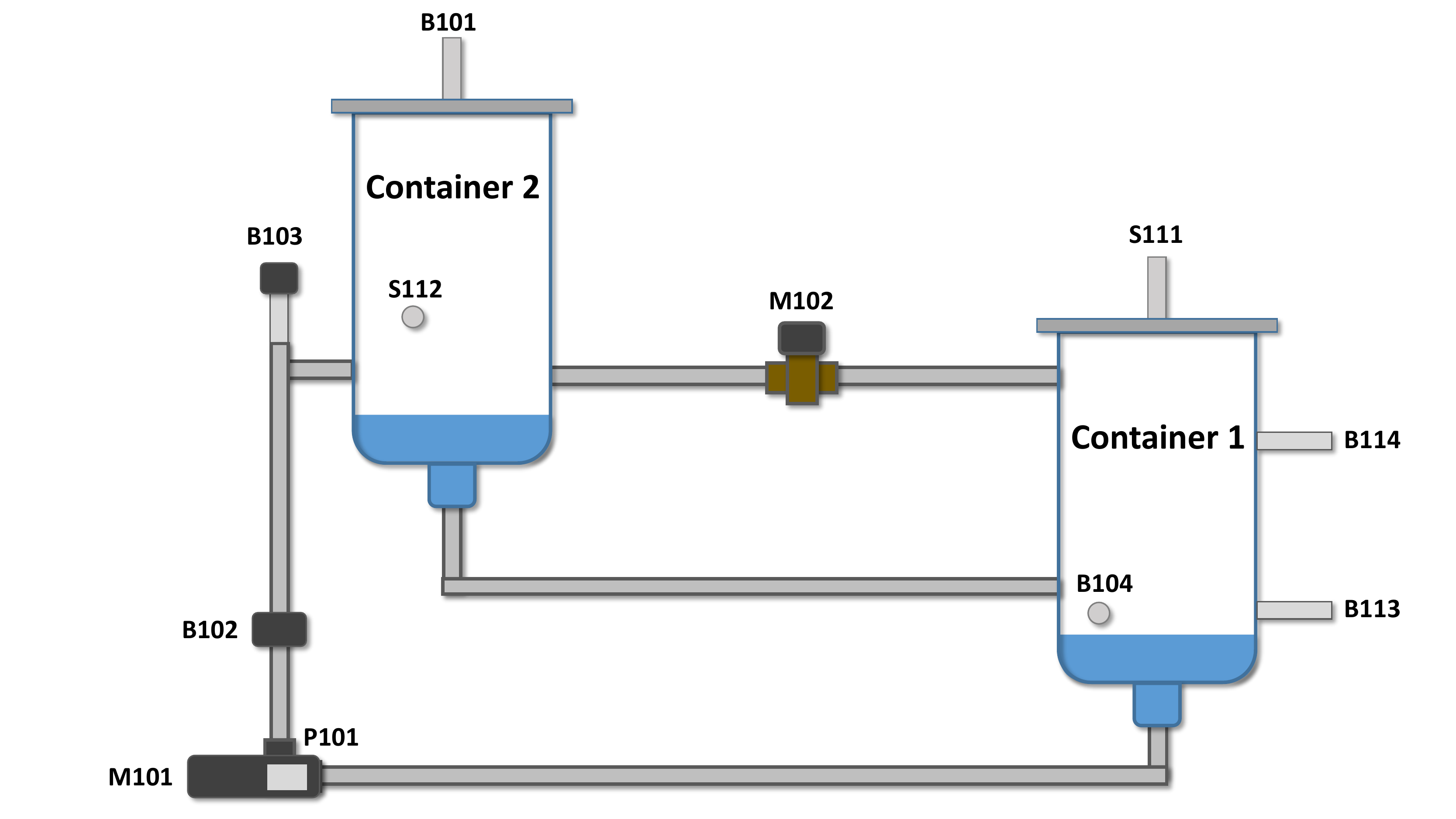}
\caption{Structure of the Process Used to Create the Data Set}
\label{fig:opcua_process}
\end{figure}

\begin{figure}[h!]
\centering
\includegraphics[width=0.48\textwidth]{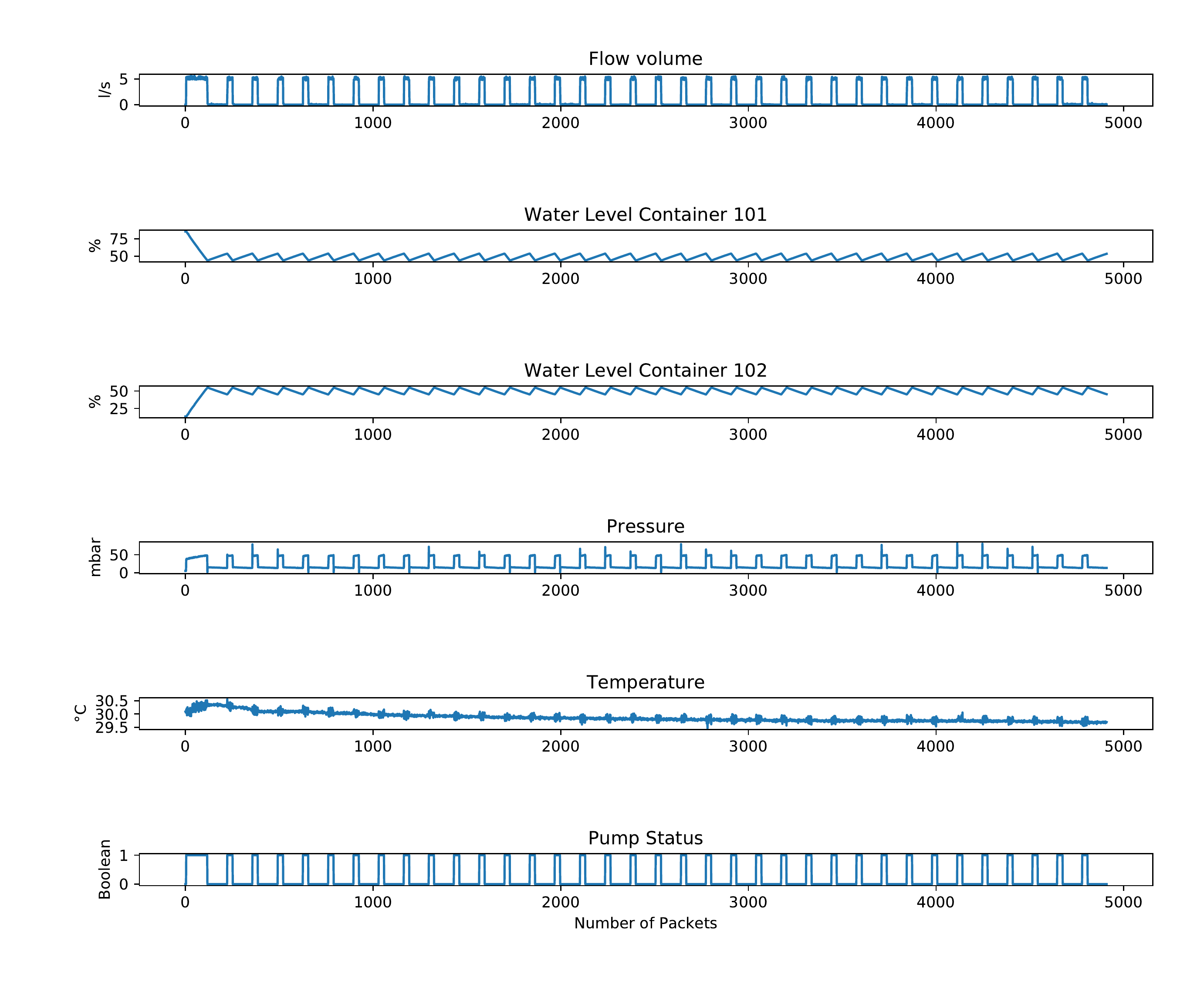}
\caption{Normal Behaviour of Process Used to Create the Data Set}
\label{fig:norm_proc_opcua}
\end{figure}

The data set is presented by \textit{Duque Anton et al.}~\cite{Duque_Anton.2019a}.
It is derived from a batch processing environment,
set up with a \textit{Festo Didactic MPS PA Compact Workstation},
a workstation for training and education purposes.
The process environment is pictured in Figure~\ref{fig:opcua_process}.
The data set is introduced by \textit{Duque Anton et al.}~\cite{Duque_Anton.2019a} and has been partially analysed by them as well~\cite{Duque_Anton.2019c}.
During normal operation,
the pump \textit{P101} transports water from \textit{Container 1} to \textit{Container 2} until an adjustable threshold value of the water level is reached.
Due to natural reflow,
the water starts to leak from \textit{Container 2} into \textit{Container 1}.
After the water level in \textit{Container 2} falls below the threshold value adjusted by a hysteresis value that is adjustable as well,
the pump starts up again and fills \textit{Container 2} again.
Normal behaviour is shown in Figure~\ref{fig:norm_proc_opcua}.
In this work,
a look is taken at the third attack scenario as described by the authors.
This scenario is derived from a process that was interrupted in two ways: 
first, 
the pump was not turned off after the threshold was reached,
leading to it running dry.
Second,
the release valve of \textit{Container 2} was opened,
leading to a constant and increased re-flow of water and more pump activity.
However,
the \ac{plc} was configured in a way that all sensors and actuator indicated normal operation.
Side-channel sensors,
such as an acoustic sensor,
were employed to still be able to detect the deviation.
The authors provided several side-channel sensor data,
of which the acoustic and flow sensor presented the most promising insights.

\section{Side-channel Sensors}
\label{sec:side_channels}
\begin{figure}[h!]
\centering
\includegraphics[width=0.48\textwidth]{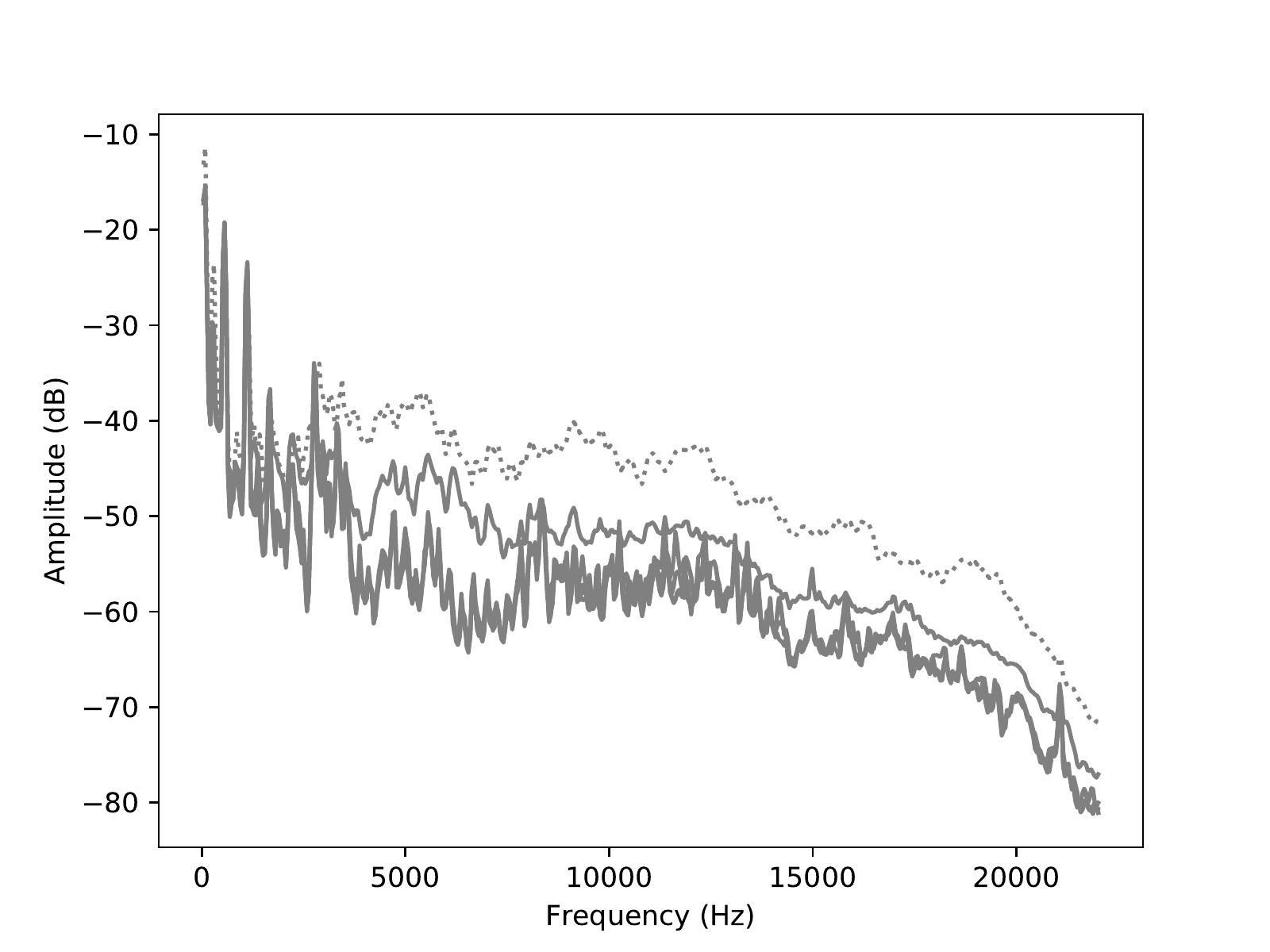}
\caption{Attack 1 in Comparison to Normal Pump Activity in the Data Set}
\label{fig:ds2_att1}
\end{figure}

\begin{figure}[h!]
\centering
\includegraphics[width=0.48\textwidth]{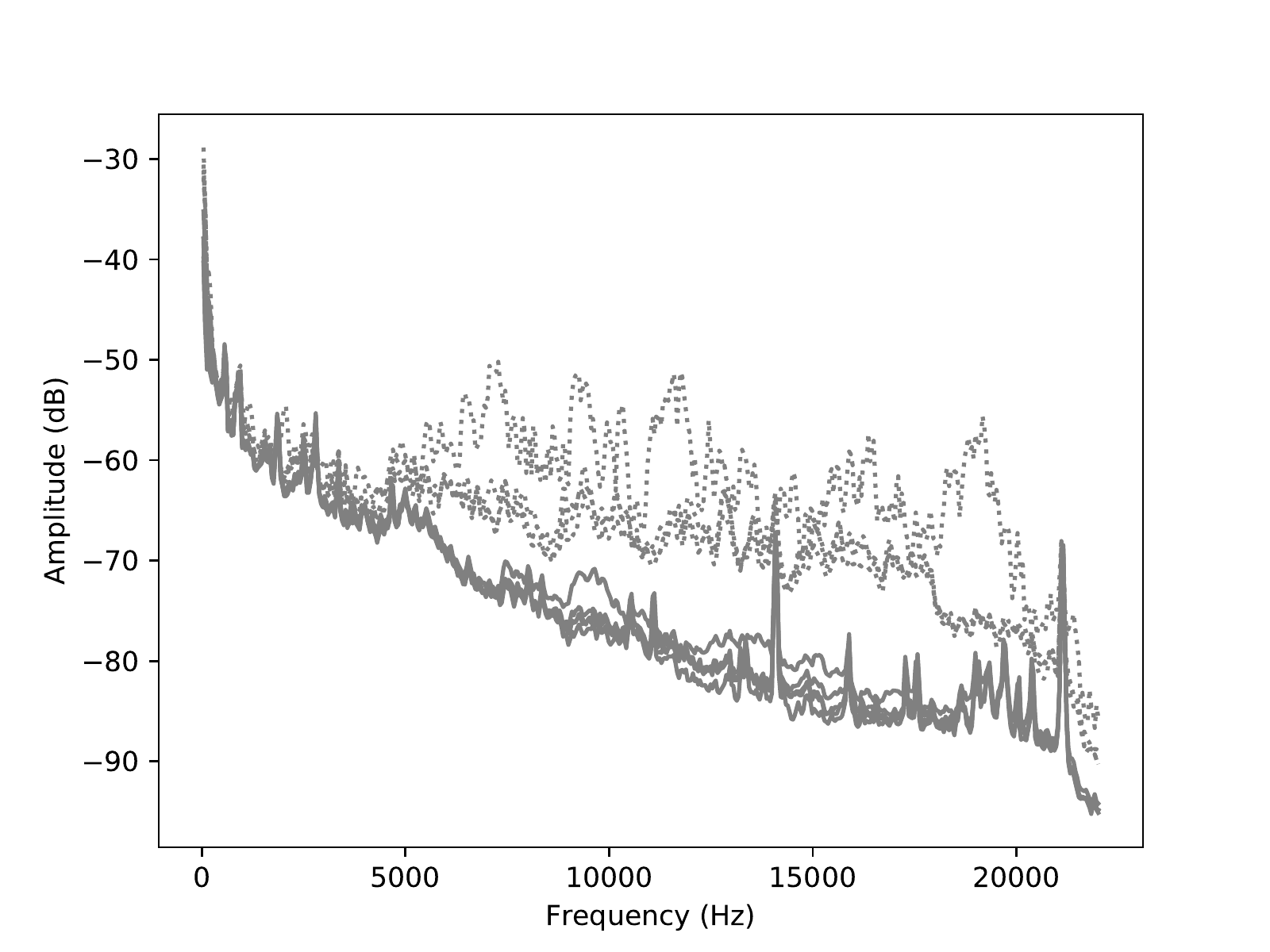}
\caption{Attack 2 in Comparison to Normal Pump Inactivity in the Data Set}
\label{fig:ds2_att2}
\end{figure}

The data set presented in Section~\ref{sec:data} covers the duration of about 20 minutes.
Thus,
the data is not sufficient for full scale machine learning-based analysis.
Instead,
a qualitative analysis is performed by hand.
Into this scenario, 
two attacks are introduced:
The pump staying active even though the threshold in Container 1 has been reached and the re-flow valve being open even though the water level is below threshold.
Since this attack provides sensor and actuator values that are not distinguishable from normal operation,
side-channel sensors are employed in order to detect attacks.
In this work,
an acoustic sensor is analysed.
In normal operation,
two different acoustic profiles can be distinguished: operation and inactivity of the pump.
The first attack changes the acoustic profile of the pump operation as shown in Figure~\ref{fig:ds2_att1}.
In this figure,
the dotted lines represent the attack behaviour,
while the solid lines were recorded during normal operation of the pump during activity.
Five events in the data set were taken for the normal data,
while there was one malicious event.
It is shown that frequencies above about \numprint{4000} contain a higher amplitude during the attack,
indicated by the dashed line.
This characteristic can be used to create an automated detection mechanism for this kind of attack based on the acoustic profile.
The second attack changes the acoustic profile while the pump is inactive,
as shown in Figure~\ref{fig:ds2_att2}.
Similar to the figure above,
the dotted lines represent the attack behaviour,
while the solid lines were recorded during normal operation of the pump during inactivity,
i.e. the pump was turned of and only re-flow of water occurred.
Five events in the data set were taken for the normal data,
while there were three malicious event.
In this case,
the amplitude of frequencies above about \numprint{5000} is significantly higher for an attack,
indicated by the dashed lines,
than during normal operation.
This feature can be used for detection as well.
In a productive environment,
profiles of different parts of the process could be created and trained.
Automated signature detection can be used to detect deviations from the normal profiles and thus detect attacks that could not be detected with conventional means.

\section{Side Channel-based Intrusion Detection}
\label{sec:sc-based_id}
As shown in the previous section,
the acoustic data is capable of distinguishing between normal and anomalous operation.
One possible metric for the detection of attacks based on anomalies in the acoustic data is extracting expressive frequencies and comparing their energy.
Promising candidates are 5,
10,
and 19 kHz.
A more sophisticated approach is the creation of motifs based on the respective spectral information,
as for example done in speech recognition,
for example with autoencoders as presented by \textit{Deng et al.}~\cite{Deng.2013}.
As the amount of motifs is expected to be relatively small due to the periodic behaviour of industrial environments,
the memory requirements are feasible.
One of the challenges,
however,
is correctly identifying the length as well as beginning and end point of the motifs,
so that different steps of processing can be assigned their corresponding motifs.
In praxis,
a microphone would monitor the acoustic properties of the device under investigation,
create a representation of the acoustic profile and compare it to the data base of known good profiles.
If there is a deviation larger than a justifiable threshold,
a human operator is alerted.
In addition to attacks,
malfunctions and the need for maintenance actions can be detected with this approach.

\section{Conclusion}
\label{sec:conclusion}
Due to the increase in attacks on industrial process environments,
intrusion detection and prevention mechanisms need to be integrated into \ac{ot} networks.
Since an abundance of infrastructure is used for long operation times that is hard to update,
industrial security solutions need to be compatible to legacy systems.
They need to integrate into existing environments.
Side channel sensors provide such means as they are relatively independent of the actual processes.
The information gathered by side-channel sensors can be used,
e.g. by industrial \ac{siem} systems,
to provide for a more holistic picture of the security stand~\cite{Duque_Anton.2019d, Duque_Anton.2019f}.
However,
processing sensor data in a meaningful fashion requires understanding of process and data,
even though automated motif extraction can aid the process of detecting or creating patterns.
In order to secure industrial environments,
new approaches need to be evaluated,
side-channels as well as emulation environments for industrial security concepts being crucial.
If such systems are combined with anomaly detection approaches,
e.g. in the timing behaviour~\cite{Duque_Anton.2018c} or on a packet basis~\cite{Duque_Anton.2018b},
attacks can be detected at several points in the industrial environment.
Deception technologies,
such as honeypots~\cite{Fraunholz.2017f},
create a defence-in-depth approach of several layered security solutions.
The future work consists of combining and implementing the individual methods into a framework for security evaluation and assessment.

%
\begin{acks}
This work has been supported by the Federal Ministry of Education and Research of the Federal Republic of Germany (Foerderkennzeichen 16KIS0932, IUNO Insec) and the Deutsche Forschungsgemeinschaft (DFG,
German Research Foundation) – 252408385 – IRTG 2057.
The authors alone are responsible for the content of the paper.
\end{acks}

%
\bibliographystyle{ACM-Reference-Format}
\bibliography{literature}

\end{document}